%% file: main.tex
\documentclass[journal]{vgtc}                     

\onlineid{1016}

\vgtccategory{Research}

\vgtcpapertype{theory/model}

\title{Too Many Cooks: Exploring How Graphical Perception Studies Influence Visualization Recommendations in Draco}

\author{%
  \authororcid{Zehua Zeng}{0000-0002-5153-3865},
  \authororcid{Junran Yang}{0000-0002-8467-2917}, \authororcid{Dominik Moritz}{0000-0002-3110-1053}, \authororcid{Jeffrey Heer}{0000-0002-6175-1655} and \authororcid{Leilani Battle}{0000-0003-3870-636X}
}

\authorfooter{
  \item
  	Zehua Zeng is with University of Maryland, College Park.
  	E-mail: zhzeng@umd.edu.
  \item
  	Junran Yang, Jeffrey Heer, and Leilani Battle are with University of Washington, Seattle.
  	E-mail: (junran, jheer, leibatt)@uw.edu.

  \item Dominik Moritz is with Carnegie Mellon University.
  	E-mail: domoritz@cmu.edu.
}

\abstract{%
  \input{content/0-abstract}
}


\graphicspath{{figs/}{figures/}{pictures/}{images/}{./}} 

\usepackage[frozencache=true,cachedir=minted-cache]{minted}
\setminted{
    fontsize=\normalsize,
    baselinestretch=1,
    breaklines=0 . 
}

\usepackage{tabu}                      
\usepackage{booktabs}                  
\usepackage{lipsum}                    
\usepackage{mwe}                       

\usepackage{mathptmx}                  

\usepackage{xspace}

\usepackage[table,svgnames,dvipsnames]{xcolor}
\usepackage{makecell}

\makeatletter
\DeclareRobustCommand\onedot{\futurelet\@let@token\@onedot}
\def\@onedot{\ifx\@let@token.\else.\null\fi\xspace}

\makeatother

\begin{document}

\renewcommand{\sectionautorefname}{Section}
\let\subsectionautorefname\sectionautorefname
\let\subsubsectionautorefname\sectionautorefname


\firstsection{Introduction}

\maketitle

\input{content/1-introduction}
\input{content/2-related-work}
\input{content/3-pipeline}
\input{content/4-case-study}
\input{content/5-analysis}

\input{content/6-discussion}

	

\acknowledgments{The authors wish to thank colleagues in the UW Interactive Data Lab, as well as Maureen Stone, Steve Franconeri, and our paper reviewers for their thoughtful feedback. This work was supported in part by the Moore Foundation and the NSF through award numbers IIS-1850115, IIS-1901386, and IIS-2141506.}

\bibliographystyle{abbrv-doi-hyperref}

\bibliography{reference}



\end{document}

%% file: content/0-abstract.tex
Findings from graphical perception can guide visualization recommendation algorithms in identifying effective visualization designs. 
However, existing algorithms use knowledge from, at best, a few studies, limiting our understanding of how complementary (or contradictory) graphical perception results influence generated recommendations. 
In this paper, we present a \textit{pipeline} of applying a large body of graphical perception results to develop new visualization recommendation algorithms and conduct an \textit{exploratory study} to investigate how results from graphical perception can alter the behavior of downstream algorithms. 
Specifically, we model graphical perception results from 30 papers in Draco---a framework to model visualization knowledge---to develop new recommendation algorithms.
By analyzing Draco-generated algorithms, we showcase the feasibility of our method to (1) identify gaps in existing graphical perception literature informing recommendation algorithms,
(2) cluster papers by their preferred design rules and constraints,
and (3) investigate why certain studies can dominate Draco's recommendations, whereas others may have little influence.
Given our findings, we discuss the potential for mutually reinforcing advancements in graphical perception and visualization recommendation research.

%% file: content/1-introduction.tex
Visualization recommendation algorithms aim to reduce time-to-insights by generating visualizations for analysts to explore with minimal or even no coding~\cite{Zeng2021evaluation,Wongsuphasawat2016towards,Zhu2020survey,Zeng2022multi-faceted}.
Visualization recommenders often leverage graphical perception guidelines and data to choose effective encodings for their recommendations.
For example, CompassQL~\cite{Wongsuphasawat2016towards} and Voyager~\cite{Wongsuphasawat2015voyager,Wongsuphasawat2017voyager2} build on Mackinlay's APT rules~\cite{Mackinlay1986automating} to refine recommendations.
Similarly, machine learning approaches rely on visualization corpora~\cite{Hu2019vizml,Li2022kg4vis} or experiment data~\cite{Moritz2018formalizing} to train models.

However, graphical perception is an extensive research space with varied theories and empirical results. 
For example, Zeng \& Battle~\cite{Zeng2023review} identify 59 graphical perception papers that could inform the development of future visualization recommendation algorithms.
Ideally, visualization recommendation algorithms would incorporate most (if not all) of these findings into the ranking process.
In contrast, current algorithms tend to incorporate findings from at most three graphical perception papers~\cite{Zhu2020survey,Zeng2023review,Saket2018beyond}.
As a result, the visualization community may be unaware of interaction effects that arise when integrating varied guidelines and empirical results in graphical perception.
How do new findings in graphical perception augment the current body of knowledge? And how do shifts within this body of knowledge translate into significant changes in downstream visualization recommendations?

To answer these questions, we must fix the disconnection between graphical perception guidelines and downstream recommendation algorithms. 
Thus, we first contribute a replicable pipeline for importing a large body of graphical perception results into visualization recommendation algorithms. Specifically, we incorporate data from  
\textit{30 different graphical perception papers} into Draco~\cite{Moritz2018formalizing,Yang2023draco2}, a visualization recommendation framework that models visualization design guidelines as a set of constraints.
To better understand the behavior of Draco-generated algorithms, we perform a case study to investigate how Draco learns visualization design knowledge with its soft constraints, including how Draco resolves both \textit{complementary} and \textit{contradictory} results.

Given the pipeline and case study, we then perform an exploratory analysis to understand (1) \textit{what} visualization design knowledge each graphical perception paper provides, and (2) \textit{how} these papers may influence downstream recommendations.
To answer the first question, we use Draco's soft constraints to represent the graphical perception knowledge gained from each paper and cluster the papers by similarities in visualization design space coverage and soft constraints.
Our analysis enables researchers to identify not only groups of graphical perception work evaluating similar visualization designs, but also visualization design decisions that may not be well covered in the literature.

To answer the second question, we train Draco with different inputs to generate multiple models: a \textit{baseline model} trained with Mackinlay's APT rules~\cite{Mackinlay1986automating}, and \textit{30 different plus-one models} trained with APT rules plus one additional graphical perception paper (derived from Zeng \& Battle's dataset~\cite{Zeng2023review}). 
We cluster graphical perception papers according to similarities in how their corresponding plus-one models induce shifts in Draco's soft constraint weights compared to the baseline.
Using the weight-shift clusters, we analyze which papers seem to share similar preferences in visualization designs, rules, and constraints.
For example, certain groups of papers seem to disagree on when to prioritize color hue encodings, even when they study similar sets of visualization designs (e.g., \cite{Jardine2020perceptual,Ondov2019face}).
We also compare our analysis results for soft constraint weights versus recommendations and confirm that soft constraint weight shifts correlate with observed shifts in visualization recommendations.

Our analysis yields a \textit{human-interpretable approach to measuring the influence of graphical perception studies}: a study's influence can be mapped to specific Draco soft constraints and, in turn, to corresponding changes in output recommendations.
Further, this approach enables \textit{automated detection of consensus and discordance among graphical perception results}, opening new avenues for meta-analysis in visualization research.
Together, these techniques contribute a novel methodology for conducting \textbf{\textit{quantitative meta-analyses}} in graphical perception. By programming this methodology to take a published dataset and schema as input~\cite{Zeng2023review} and output results using an established framework~\cite{Moritz2018formalizing}, our results are \textit{transparent, replicable, and extendable} by the visualization community. Further, these methods have the potential to generalize beyond Draco by augmenting the output targets in our pipeline, e.g., from Draco constraints to queries in other visualization recommendation frameworks~\cite{Wongsuphasawat2016towards,Siddiqui2017fast}.

Our analysis also reveals specific blind spots within Draco's knowledge base, suggesting new opportunities to \textit{extend existing visualization recommendation frameworks} to cover the latest contributions in graphical perception.
Finally, we acknowledge that our work is \textit{exploratory}. Our observations depend on how graphical perception results are interpreted and how existing visualization recommendation frameworks can model them. Despite these limitations, we believe our analysis provides a unique perspective on how research in graphical perception can be analyzed and disseminated more broadly.

We also share all our data and code to facilitate reuse: \url{https://github.com/Zehua-Zeng/too-many-cooks}

%% file: content/2-related-work.tex
\vspace{-1mm}
\section{Related Work}

In this paper, we present an exploratory study to understand how the behaviors and outputs of visualization recommendation algorithms change when trained on different sets of graphical perception results.
In this section, we discuss existing graphical perception research and how current visualization recommendation algorithms use these results. Then, we discuss why, given several existing visualization recommendation frameworks, we choose Draco~\cite{Moritz2018formalizing, Yang2023draco2} for modeling a large body of graphical perception results.

\vspace{-2mm}
\subsection{Graphical Perception Knowledge}

The design of ``effective'' visualizations is a long-standing topic of study~\cite{Zeng2023review}, resulting in a panoply of theoretical guidelines to select appropriate visual encodings based on data types and analysis tasks~\cite{Bertin1983semiology,Mackinlay1986automating,Cleveland1984graphical,Shneiderman1996eye,Mackinlay2007showme}.
Experimental work aims to validate these theoretical rules through user studies~\cite{Zeng2023review,Ware2012information}.
For example, Cleveland \& McGill~\cite{Cleveland1984graphical} verified part of their hypothesized encoding ranking by running experiments to compare the effectiveness of conveying quantitative values between position and length encodings, as well as position and angle encodings. 
Later, their experimental results were replicated and extended by Heer \& Bostock's~\cite{Heer2010crowdsourcing} crowdsourcing graphical perception experiments.
Other experiments, such as those conducted by Kim \& Heer~\cite{Kim2018assessing} and Saket et al.~\cite{Saket2018task}, evaluate the performance of many visualization designs for a specific set of visual analysis tasks.

Existing surveys of graphical perception work only summarize visualization design guidelines for visualization designers~\cite{Ware2012information,Quadri2022survey}, which do not contribute datasets that can be directly imported into visualization recommendation algorithms.
Because of the lack of consistent and shared datasets across studies, only a few of these graphical perception results have been applied in downstream recommendation algorithms~\cite{Saket2018beyond,Zhu2020survey}, restricting the visualization community's ability to reason about potential synergies or conflicts among perception guidelines as a whole.
To resolve this problem, Zeng \& Battle~\cite{Zeng2023review} presented a survey to identify the graphical perception literature most relevant to visualization recommendation and contributed the \textit{first dataset} of graphical perception results that can be easily imported into existing systems.
However, we have yet to see any \textit{practical applications} of the Zeng \& Battle dataset in visualization recommendation work.
In this paper, we take one step further to investigate the feasibility of applying a large body of graphical perception work to visualization recommendation algorithms and analyze how guidelines and findings from different papers alter the behavior of visualization recommendation algorithms.

\vspace{-2mm}
\subsection{Visualization Recommendation Algorithms}

Visualization recommendation algorithms rely on results from graphical perception to compare and rank visualizations of potential interest, such as empirical comparisons or theoretical orderings of encodings under certain perceptual tasks~\cite{Zhu2020survey,Zeng2023review,Shen2021towards}.
Unlike machine-learning recommendation algorithms which require extensive human-labeled corpora to train models~\cite{Li2022kg4vis,Moritz2018formalizing,Hu2019vizml}, rule-based algorithms use existing theoretical guidelines~\cite{Wongsuphasawat2015voyager,Wongsuphasawat2017voyager2,Mackinlay2007showme} or propose new ranking metrics~\cite{Demiralp2017foresight,Vartak2015seedb,Key2012vizdeck} to compare candidate visualizations.
For example, Wongsuphasawat et al.~\cite{Wongsuphasawat2015voyager,Wongsuphasawat2017voyager2} use Mackinlay's expressiveness and effectiveness principles~\cite{Mackinlay1986automating} to produce recommendations.
Vartak et al.~\cite{Vartak2015seedb} propose an ``interestingness'' metric based on deviations in the data. Key et al.~\cite{Key2012vizdeck} and Demiralp et al.~\cite{Demiralp2017foresight} use statistical features of the dataset to approximate potential insights to guide exploratory visual analysis.

All of these visualization recommendation algorithms use different metrics to determine which visualizations to suggest to the user~\cite{Zeng2021evaluation,Zeng2022multi-faceted}. However, few graphical perception results are applied in existing algorithms~\cite{Zeng2023review}, calling into question the veracity of their metrics with respect to established perceptual guidelines.
Incorporating more graphical perception data and guidelines should presumably affect an algorithm's behavior.
However, existing research has yet to investigate how a large body of graphical perception results influences a recommendation algorithm's behavior and so, by extension, cannot quantify the unique contributions of individual papers on an algorithm's output.

\vspace{-2mm}
\subsection{Visualization Recommendation Frameworks}

Several frameworks~\cite{Moritz2018formalizing,Wongsuphasawat2016towards,Siddiqui2017fast} have been proposed to streamline the development of new visualization recommendation algorithms.
Although one can use CompassQL~\cite{Wongsuphasawat2016towards} and ZQL~\cite{Siddiqui2017fast} to generate new algorithms,
they lack the infrastructure required to import graphical perception results as input data to generate visualization recommendations. In contrast, Draco~\cite{Moritz2018formalizing,Yang2023draco2} supports modeling both theoretical rules and experimental results as a set of \textit{constraints} on the visualization design space. By assigning weights to constraints, Draco can also prioritize certain sub-spaces within the enumerable visualization design space to efficiently rank visualization recommendations.

Rule-based recommendation algorithms can technically apply theoretical or experimental perception results to guide visualization recommendation, but the integration of these results is performed \textit{manually} by the algorithm designer with hand-tuning.
Compared to other frameworks, Draco is the first (and only) one to \textit{automate} the integration of graphical perception results (both theoretical and empirical) in the design of new visualization recommendation algorithms. 
To the best of our knowledge, this paper is the first to investigate how incorporating different amounts and combinations of graphical perception results into Draco affects the behavior of the resulting recommendations. 

%% file: content/3-pipeline.tex
\vspace{-1mm}
\section{Pipeline: Modeling Existing Graphical Perception Results in Draco}
\label{sec:pipeline}

In this section, we explain how we constructed a pipeline for modeling graphical perception results collated by Zeng \& Battle~\cite{Zeng2023review} in Draco~\cite{Yang2023draco2,Moritz2018formalizing}.
To help readers better understand the inputs and outputs of the pipeline, we briefly summarize the contributions and limitations of Zeng \& Battle's survey~\cite{Zeng2023review} and Draco~\cite{Yang2023draco2,Moritz2018formalizing}.

\vspace{-1mm}
\subsection{Zeng \& Battle's Survey: Collation of Existing Graphical Perception Studies}
\label{sec:pipeline:zeng-battle}

\begin{figure}
\centering
 \includegraphics[width=0.90\columnwidth]{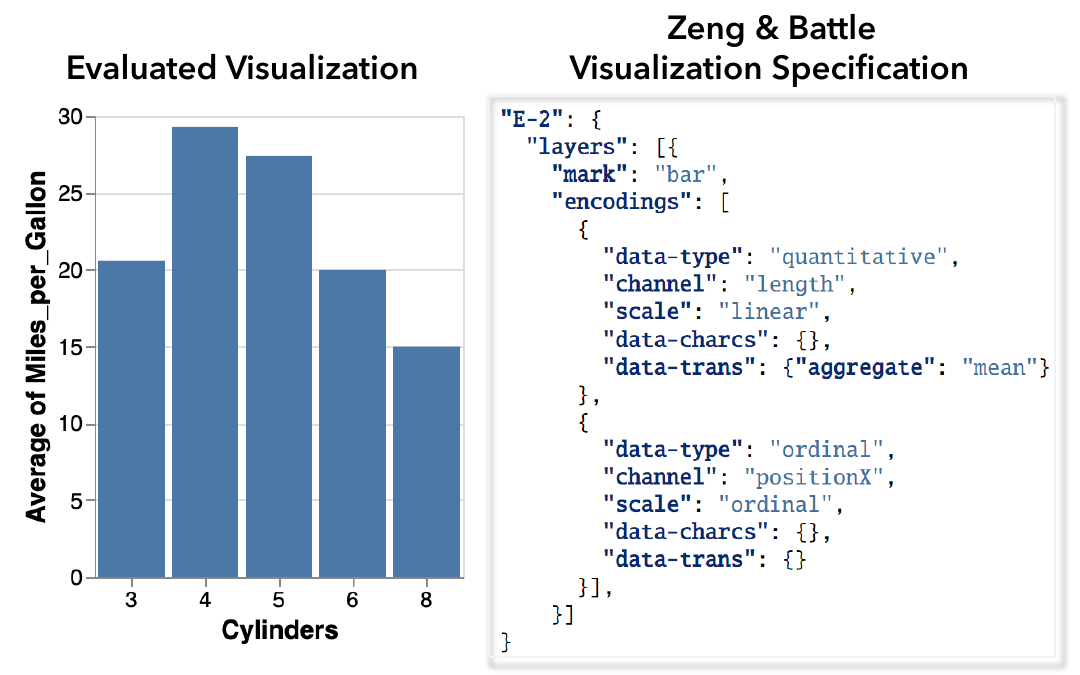}
 \caption{The visualization specification used in Zeng \& Battle's dataset.}
 \label{fig:zeng-battle-exp}
 \vspace{-5mm}
\end{figure}

In this paper, we rely on data published by Zeng \& Battle~\cite{Zeng2023review}, where they collated 59 different graphical perception papers into a consistent, JSON-based dataset that can be imported into visualization recommendation algorithms. 
Each paper in this dataset reports a series of theories or experiments where \textit{visualization designs} are compared under one or more \textit{graphical perception tasks}. 
To record evaluated visualization designs, Zeng \& Battle extended the Vega-Lite specification~\cite{Satyanarayan2016vegalite} to capture \textit{data types, data characteristics, data transformations, encodings, marks,} and \textit{scales} (as shown in~\autoref{fig:zeng-battle-exp}).
To document the graphical perception results, Zeng \& Battle recorded the performance rankings of evaluated visualization pairings (with \textit{statistical significance} and \textit{effect sizes}) under specific \textit{graphical perception tasks} and \textit{metrics}.

\subsubsection{Limitations}

Note that current visualization recommendation algorithms tend to omit specific visualization types, which by extension are also excluded from Zeng \& Battle's survey, such as 3D, graph, animated, and interactive visualizations.
Also, although Zeng \& Battle extended Vega-Lite to accommodate more graphical perception studies, their extension (in~\autoref{fig:zeng-battle-exp}) does not yet capture more granular design decisions, for example, how the bars are sorted~\cite{Zhao2019neighborhood} or whether distractors are added~\cite{Talbot2014four}, etc.
Moreover, general information of graphical perception studies, like the sample size of graphical perception studies and the demographic background of participants, etc., are not captured in Zeng \& Battle's dataset. Still, this survey is the first to provide a comprehensive corpus of graphical perception knowledge for direct import by visualization recommendation systems, hence why we use it in this paper.

\subsection{Draco: Modeling Visualization Design Knowledge}
\label{sec:pipeline:draco}

Draco~\cite{Moritz2018formalizing,Yang2023draco2} uses a collection of hard and soft constraints to represent visualization design knowledge as guidelines for generating visualization recommendations.
Hard constraints shape the visualization design space by eliminating ill-formed or non-expressive designs from all possible combinations; thus, a visualization must satisfy all hard constraints to be considered valid.
For example, the hard constraint 
\vspace{-2mm}
\begin{minted}{prolog}
violation(log_non_positive) :-
attribute((field,min),F,MIN),
helper((encoding,field),E,F),
helper((encoding,scale_type),E,log),
MIN <= 0.
\end{minted}
\vspace{-2mm}
tells Draco to eliminate visualization designs that apply a log scale to a data attribute containing zero or negative values.
In contrast, soft constraints represent visualization \textit{preferences}; they can be violated with a corresponding weighted cost.
For example, the soft constraint 
\vspace{-2mm}
\begin{minted}{prolog}
#const high_cardinality_shape_weight = 10.
preference(high_cardinality_shape, E) :-
helper(encoding_cardinality,E,N),
attribute((encoding,channel),E,shape),
N > 8.
\end{minted}
\vspace{-2mm}
states that for any encoding \texttt{E}, using the \texttt{shape} channel to represent high-cardinality data incurs a cost of \texttt{10}. 
The \textit{Draco cost} of a visualization is the sum of the costs of all violated soft constraints.
Given an input query, Draco uses the Clingo solver~\cite{Gebser2011potassco,Gebser2014clingo} to search the visualization design space, screening valid visualizations via hard constraints and nominating designs with the \textbf{\textit{lowest}} Draco soft constraint costs.

There are two ways to obtain soft constraint weights.
First, Draco allows algorithm designers to define their own soft constraints and manually assign a weight to each constraint to indicate their preferences.
Second, Draco-Learn can learn weights for existing soft constraints from ranked pairs of visualizations.
For example, multiple studies~\cite{Cleveland1984graphical,Heer2010crowdsourcing,Waldner2019comparison} find that people tend to be better at making quantitative comparisons using bar charts than pie/radial charts. The comparisons made in these experiments can be translated into ranked pairings of visualizations recognizable by Draco-Learn.
A similar strategy can be applied to harvest rankings from theoretical papers as well.

\subsubsection{Limitations}
\label{sec:pipeline:draco-limitations}

Note that Draco is still under active development~\cite{Yang2023draco2}.
In this paper, we use a recent version of 
Draco\footnote{\small\url{https://github.com/cmudig/Draco/tree/023e9e}} with a similar architecture but a cleaner API, more tests, and a more flexible chart specification format, compared to the original (\cite{Moritz2018formalizing}). 
Still, the newer Draco has several limitations in its corresponding Draco-Learn application.
First, Draco's visualization design space is narrower than the Zeng \& Battle and Vega-Lite specifications.
For example, visualization types such as maps, parallel coordinates, etc., cannot be represented by the Draco specification.
Moreover, some granular design decisions cannot be captured by Draco although they are recorded in Zeng \& Battle's dataset, e.g., custom color or shape palettes. 
Second, Zeng \& Battle observed ten types of visual analysis tasks across graphical perception studies; however, Draco can only differentiate between two task categories: \texttt{summary} and \texttt{value}~\cite{Kim2018assessing,Zeng2023review}. 
Third, although Draco-Learn can learn weights from ranked pairs of visualizations, it does not take the statistical test results into account, such as the statistical method, threshold, or effect size, even though they are recorded in Zeng \& Battle's dataset.

\subsection{Incorporating Zeng \& Battle's Dataset into Draco}
\label{sec:pipeline:incorporate-dataset-to-draco}

\begin{figure*}
\centering
 \includegraphics[width=1.0\textwidth]{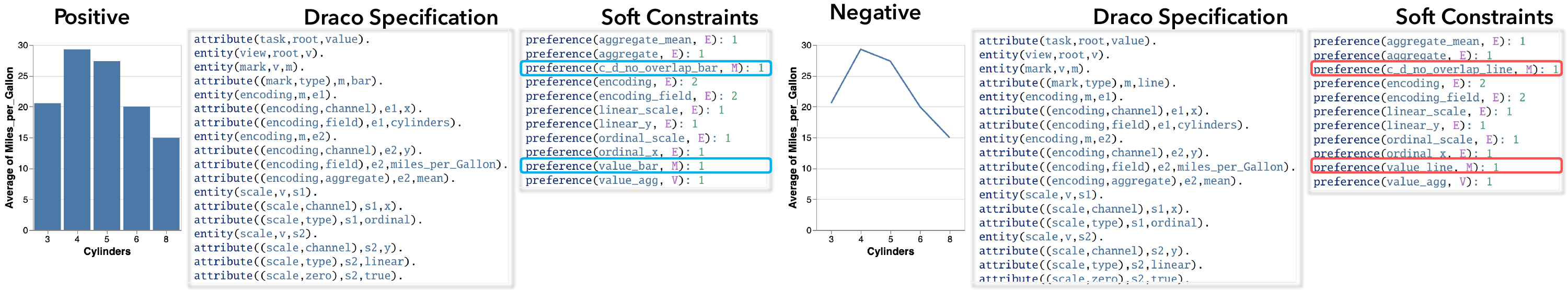}
 \caption{A demonstration of the Draco visualization specification and how Draco-Learn can detect and then learn visualization preferences from a ranked visualization pair concluded from \texttt{Saket2018task}~\cite{Saket2018task}.}
 \label{fig:saket-draco-example}
 \vspace{-5mm}
\end{figure*}

After discussing the contributions and limitations of Zeng \& Battle's survey~\cite{Zeng2023review} and Draco~\cite{Moritz2018formalizing,Yang2023draco2}, we now describe how we construct our pipeline for modeling both theoretical and experimental visualization comparisons as Draco constraints.

\subsubsection{Learning from Experimental Results}

As mentioned in~\autoref{sec:pipeline:draco}, Draco-Learn can learn soft constraint weights from pairs of ranked visualizations, where the positive example performs better than the negative example.
Many empirical studies, e.g., \cite{Cleveland1984graphical,Heer2010crowdsourcing,Kim2018assessing,Saket2018task}, already evaluate various visualization designs under different analysis tasks.
Here, we use one ranked pair of visualizations tested by Saket et al.~\cite{Saket2018task} (labeled as \texttt{Saket2018task}, see~\autoref{fig:saket-draco-example}) to demonstrate how we translate experimental results from such pairs into training data for Draco-Learn.
Specifically, \texttt{Saket2018task} find that ``bar charts perform significantly better than line charts in terms of \texttt{accuracy} with the \texttt{Retrieve Value} task''. 
To create ranked pairs, we wrote a program for automatically translating Zeng \& Battle's visualization specification (\autoref{fig:zeng-battle-exp}) to Draco's (\autoref{fig:saket-draco-example}). 

When passed this pair of visualizations as input, Draco-Learn counts the number of constraint violations from the positive and negative visualizations and generates corresponding feature vectors. Draco-Learn then learns the preference differences of the ranked pair.
For example, ``\mintinline{prolog}{preference(encoding,E): 2}'' states that the visualization contains ``two'' encoding channels.
As shown in~\autoref{fig:saket-draco-example}, 
Draco-Learn detects soft constraints \mintinline{prolog}{preference(c_d_no_overlap_bar, E)} and \mintinline{prolog}{preference(value_bar,E)} (highlighted in blue) in the positive example (bar chart), which are not seen in the negative example (line chart).
Meanwhile, Draco-Learn detects that soft constraints \mintinline{prolog}{preference(c_d_no_overlap_line, E)} and \mintinline{prolog}{preference(value_line, E)} (highlighted in red) are only seen in the negative example.
Learning from this ranked visualization pair, Draco potentially would prefer (i.e., decrease the weights of) the two soft constraints that are detected only in the positive example, and disfavor (i.e., increase the weights of) the two soft constraints seen only in the negative example. 

\subsubsection{Learning from Theoretical Principles}

\begin{figure*}
\centering
 \includegraphics[width=1.0\textwidth]{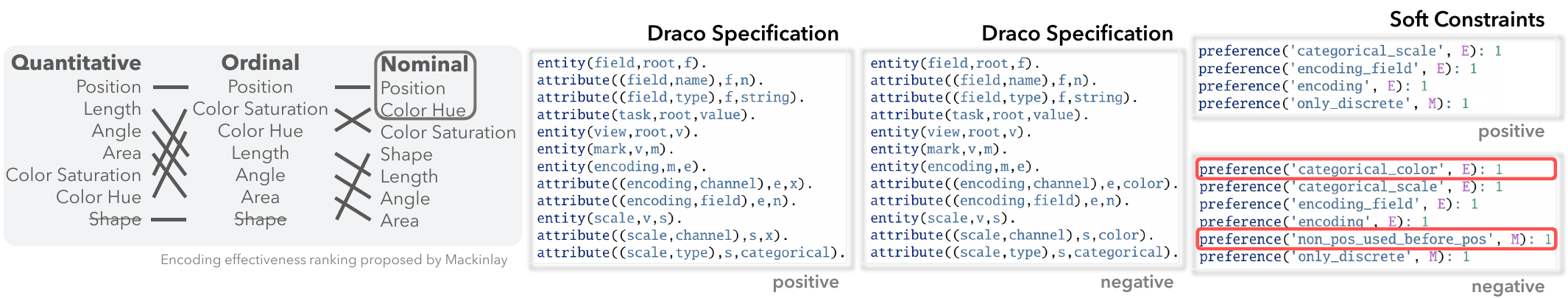}
 \vspace{-2mm}
 \caption{A demonstration of translating existing theoretical perception rules~\cite{Mackinlay1986automating} to Draco-Learn~\cite{Yang2023draco2} training data. When comparing specifications encoding nominal data using position-x (positive) versus color hue (negative), Draco learns to avoid categorical color hue encodings (in red). }
 \label{fig:draco-model-apt}
 \vspace{-5mm}
\end{figure*}

Moritz et al.~\cite{Moritz2018formalizing} use visualization pairs to teach empirical findings to Draco-Learn, but use manual weight assignment to account for existing theoretical perception rules.
Draco-APT and Draco-CQL are examples of how they manually implemented preference rules from two theoretical hypotheses, APT~\cite{Mackinlay1986automating} and CompassQL~\cite{Wongsuphasawat2016towards}, into two sets of soft constraints independently.
However, it is unclear how to measure the impact of hand-tuned weights on Draco's behavior compared to weights derived by Draco-Learn. Thus, we translate existing theoretical rankings into corresponding visualization pairs and use Draco-Learn to infer soft constraint weights. In this way, we can generate consistent data (visualization pairings) for both result types from Zeng \& Battle's dataset and can quantify their influence on Draco's behavior.

APT rules~\cite{Mackinlay1986automating} provide an encoding ranking based on its effectiveness in visualizing each high-level data type (quantitative, ordinal, and nominal).
We use one of the APT rules (\textit{prefer to use position encodings over color-hue encoding for nominal data}) to demonstrate how to translate theoretical rules into input visualization pairings for Draco-Learn (\autoref{fig:draco-model-apt}).
\texttt{Nominal} data type is implied by the combination of the field type (string) and the scale type (categorical). From~\autoref{fig:draco-model-apt}, we can see that the applied encoding channel is the only difference between the positive and negative visualization specifications.

\subsubsection{Pipeline Summary}

We only extract the statistically significant pairs of visualizations from Zeng \& Battle's dataset and contribute programs to translate them into equivalent Draco specifications. 
To keep our analysis process realistic and consistent, we chose not to modify any of Draco's default soft constraints.
We observe that some visualization designs could not be represented (due to the limitations of Draco, see~\autoref{sec:pipeline:draco-limitations}), for example, distinguishing visualizations using shape encodings with different sets of shapes~\cite{Burlinson2018open}. 
As a result, we could only translate a subset of Zeng \& Battle's corpus into Draco soft constraints, revealing exciting avenues for future work in developing robust frameworks for visualization recommendation (which we discuss in \autoref{subsec:discussion:benefit-framework}). That being said, we \textbf{successfully translate 30 papers from Zeng \& Battle's corpus into Draco visualization specifications}, affording a larger scale of quantitative meta-analysis than seen in prior work~\cite{Moritz2018formalizing}.

%% file: content/4-case-study.tex
\section{Case Study: How Does Draco-Learn Shift Soft Constraint Weights?}
\label{sec:draco-case-study}

\begin{figure*}
\centering
 \includegraphics[width=1.0\textwidth]{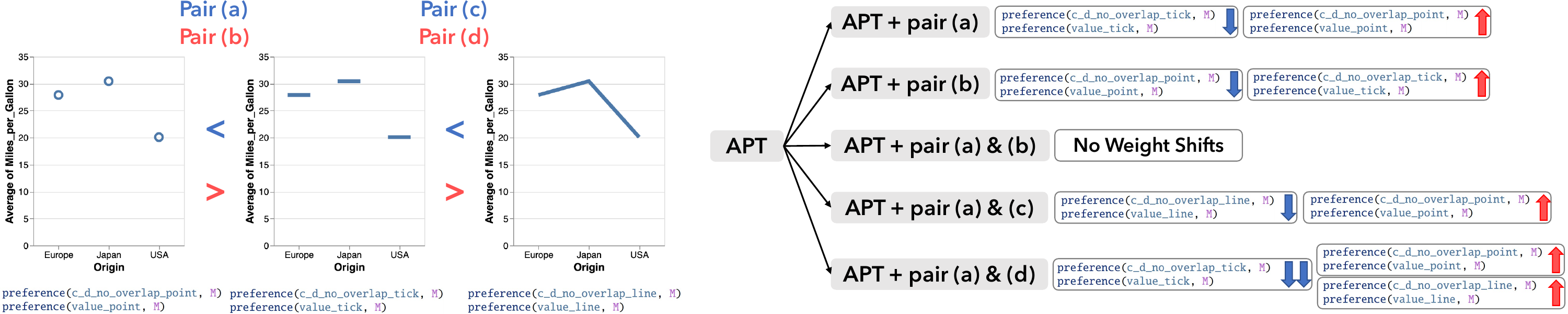}
 \caption{We conduct mini-experiments to investigate how Draco-Learn shifts the weights of soft constraints based on different training data. The left side shows the four ranked visualization pairs applied in the mini-experiments, as well as the soft constraints that are only seen in the corresponding visualization. The right side shows how the weights of soft constraints shift compared to the baseline for each mini-experiment. \vspace{-5mm}}
 \label{fig:demo-draco}
\end{figure*}

Theoretically, Draco-Learn would assign lower costs to soft constraints in the positive design choices and higher costs to soft constraints in the negative ones.
However, Draco-Learn still remains as a black box since none of the Draco publications~\cite{Moritz2018formalizing,Yang2023draco2} has demonstrated how Draco reconciles agreements and disagreements across multiple result sets.
Thus, we conduct a series of mini-experiments to test the possible scenarios. We observe that any potential interaction effects among result sets must fall into at least one of the following five categories: they fully agree, partially agree, partially disagree, fully disagree, or do not overlap. In the case of non-overlap, Draco essentially unions the orthogonal result sets; thus, we omit this scenario from our mini-experiments. For the remaining scenarios, we share the results of our mini-experiments, which demonstrate how Draco resolves them on a carefully controlled set of visualization pairings.

\textbf{Mini-Experiment Setup.} All of our mini-experiments draw from the same set of ranked visualization pairs, shown in~\autoref{fig:demo-draco}.
All visualizations within these pairs visualize the same dataset using the same encodings; they only differ in their preferred mark types:
\begin{itemize}[nosep]
    \item Pair (a): prefers tick marks over point marks
    \item Pair (b): prefers point marks over tick marks
    \item Pair (c): prefers line marks over tick marks
    \item Pair (d): prefers tick marks over line marks
\end{itemize}

Note that Draco-Learn, as a machine-learning method, expects a \textit{corpus} to be provided for training purposes rather than just one data point~\cite{Moritz2018formalizing,Herbrich2000large}.
Moreover, to analyze how soft constraint weights shift after adding specific visualization pair(s), we need a ``baseline'' to compare with.
We select the APT corpus~\cite{Zeng2023review, Mackinlay1986automating} to train a baseline model (Draco-APT) for three reasons:
(1) APT is a fundamental and influential work in graphical perception, (2) it provides a sufficient number of ranked visualization pairs for training Draco-Learn,
and (3) it is a core use case from the original Draco paper~\cite{Moritz2018formalizing}.
For each mini-experiment, we compare our baseline APT model with an \textit{``experimental''} model, which uses the APT corpus and a specific subset of our target visualization pairs ((a) through (d) above) as a training dataset.
Then, we compare the soft constraint weights of the baseline and experimental models to investigate how Draco-Learn shifts soft constraint weights under the corresponding agreement/disagreement scenarios.

\vspace{-1.5mm}
\subsection{Exp A. Adding One Single Pair (a)}

As shown in~\autoref{fig:demo-draco}, \mintinline{prolog}{preference(c_d_no_overlap_tick, E)} and \mintinline{prolog}{preference(value_tick, E)} are only seen in the positive example of pair (a), and \mintinline{prolog}{preference(c_d_no_overlap_point, E)} and \mintinline{prolog}{preference(value_point, E)} only in the negative example.
Comparing the soft constraint weights between the baseline and experimental models, Draco-Learn decreases the weights for \mintinline{prolog}{preference(c_d_no_overlap_tick, E)} and \mintinline{prolog}{preference(value_tick, E)} but increases the weights for \mintinline{prolog}{preference(c_d_no_overlap_point, E)} and \mintinline{prolog}{preference(value_point, E)} in the experimental model.
Given that Draco recommends visualizations with \textit{minimum cost}, a decrease in weight indicates that Draco will prioritize visualizations matching this soft constraint.
To summarize, adding only (a) to our baseline model causes Draco to \textit{prefer the new positive example} and shift its soft constraint weights accordingly.

\vspace{-1mm}
\subsection{Exp B. Adding One Single Pair (b)}

Pair (b) is the exact opposite of pair (a): it prefers point marks over tick marks. 
Compared to the baseline, the experimental model decreases the weights for \mintinline{prolog}{preference(c_d_no_overlap_point, E)} and \mintinline{prolog}{preference(value_point, E)} but increases the weights for \mintinline{prolog}{preference(c_d_no_overlap_tick, E)} and \mintinline{prolog}{preference(value_tick, E)}, which is the opposite of the result of Exp A. These findings are consistent with the idea that for a  single pairing added, Draco will prefer the new positive example and shift its weights accordingly.

\vspace{-1mm}
\subsection{Exp C. Adding Completely Conflicting Pairs (a) \& (b)}

After inputting both pairs (a) and (b) into the training dataset, we find that the experimental model has the exact same soft constraint weights as the baseline model.
This result indicates that by inputting completely conflicting visualization comparison knowledge into Draco-Learn, Draco-Learn will have zero weight shifts in the soft constraints.
In other words, \textit{conflicting result sets with perfect overlap may completely cancel each other out in Draco}.

\vspace{-1mm}
\subsection{Exp D. Adding Partially Conflicting Pairs (a) \& (c)}

Pairs (a) and (c) partially conflict with each other. 
The visualization with tick marks is the positive example in pair (a) but the negative example in pair (c). Otherwise, (a) and (c) do not overlap.
After inputting both pairs into Draco-Learn, we find that the weights of tick-related constraints \mintinline{prolog}{preference(c_d_no_overlap_tick, E)} and \mintinline{prolog}{preference(value_tick, E)} are still the same between the baseline and experimental models.
However, the weights for point-related constraints \mintinline{prolog}{preference(c_d_no_overlap_point, E)} and \mintinline{prolog}{preference(value_point, E)} are increased, and the weights for line-related constraints \mintinline{prolog}{preference(c_d_no_overlap_line, E)} and \mintinline{prolog}{preference(value_line, E)} are decreased.
To summarize, when Draco encounters partially conflicting results, it seems to \textit{cancel out conflicting overlaps and union the difference} between result sets. Note that this scenario appears to decompose nicely into a combination of our ``fully disagree'' (Exp C) and ``do not overlap'' scenarios, suggesting that Draco is performing straightforward calculations on the unions and differences between observed result sets. We test whether this assumption holds in our subsequent mini-experiments.

\vspace{-1mm}
\subsection{Exp E. Adding Partially Agreeing Pairs (a) \& (d)}

Pairs (a) and (d) partially agree with each other. 
They both prefer using tick marks, but pair (a) penalizes point marks while pair (d) penalizes line marks.
We find that the corresponding experimental model decreases the weights of tick-related constraints and increases the weights of the point-related and line-related constraints.
Furthermore, we observe that the negative weights for tick-related constraints appear to be double the magnitude of the positive weights for the point- and line-related constraints.
This result seems to indicate that aligned overlaps have an additive effect on soft constraint weights in Draco. In other words, Draco seems to \textit{add aligned overlaps together as well as union the difference} between result sets. Furthermore, these results seem to corroborate the idea that Draco handles partial overlap as a union of the complete overlap and complete non-overlap scenarios.

\vspace{-1mm}
\subsection{Exp F. Adding Duplicate Single Pairs (a)}

From Exp E, we find that if a soft constraint is detected on one side (positive or negative example) more than one time, Draco-Learn tends to shift its weight more than other constraints that are only seen once.
Although the results of Exp E and Exp D were additive, it is also possible that Draco does not assume a perfect additive or subtractive effect between overlapping result sets. To test this theory, we repeat Exp A and observe how Draco handles varying numbers of duplicate pairs (specifically, pair (a)).

\begin{figure}
\centering
 \includegraphics[width=0.8\columnwidth]{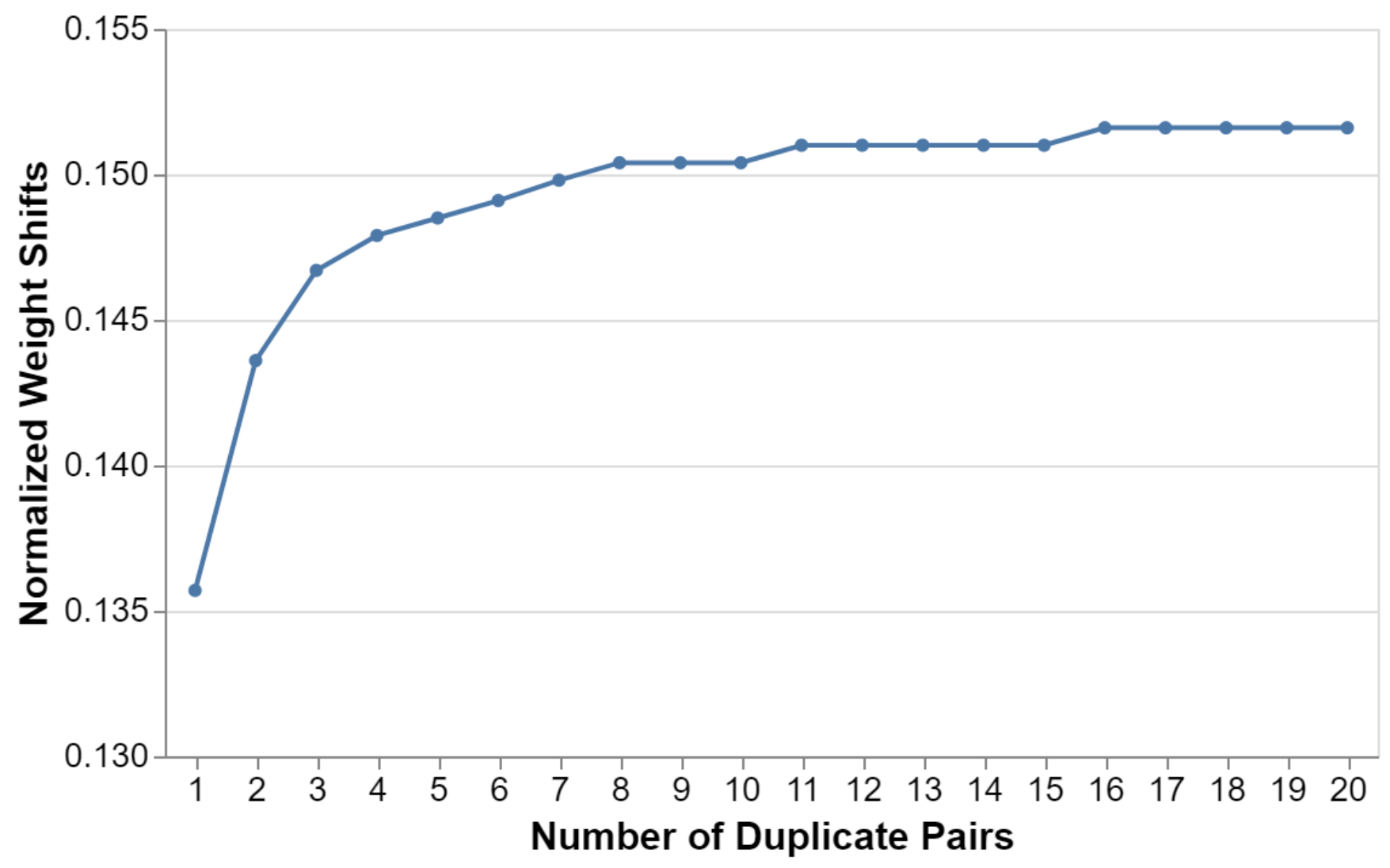}
 \vspace{-2mm}
 \caption{The trend in weight shifts when adding duplicate single pairs. Including pair (a) multiple times increases the weights of the corresponding constraints almost linearly at first but levels out after about five inclusions.}
 \label{fig:draco-duplicate-pairs}
 \vspace{-5mm}
\end{figure}

Note that Draco-Learn might assign weights to soft constraints with different scales in different runs (with different training datasets)~\cite{Yang2023draco2}.
As a result, the \textit{raw values} of soft constraint weights can vary considerably, up to the thousands.
Thus, instead of analyzing the absolute value of weight shifts, we normalize the weight shifts:
\vspace{-2mm}
\begin{equation}
    n_i = (e_i - b_i) / e_{max};
    \enspace where \enspace e_{max} = \max (|e_1|, |e_2|,.., |e_n|)
    \vspace{-2mm}
\end{equation}
Here, $b_i$ is the weight of soft constraint $i$ under the baseline model, $e_i$ is the weight of soft constraint $i$ under the experimental model, and $e_{max}$ is the maximum absolute weight of soft constraints among all soft constraints under the experimental model.
$n_i$ is the resulting normalized weight shift for soft constraint $i$.
Since all four constraints (tick-/point-related) shifted the same absolute amount in the same experimental run, we only draw one weight shift trend in~\autoref{fig:draco-duplicate-pairs}.
From~\autoref{fig:draco-duplicate-pairs}, we can see that duplicate visualization pairings can shift the related weights more than pairings only observed once, but this effect levels off with increasing numbers of duplicates.
To summarize, Draco will emphasize pairings it has seen multiple times, but we see diminishing returns, i.e., seeing 2 copies of a pairing may double the weight(s) but 20 copies likely will not translate to a 20$\times$ change in weight(s).

\vspace{-2mm}
\subsection{Case Study Summary}

To summarize, we make the following observations from our mini-experiments, shown in \autoref{fig:demo-draco}: (1) Draco seems to decompose partial overlaps into a union of complete overlap and non-overlap scenarios (Exp D, Exp E, Exp F), (2) complete disagreement seems to cause soft constraint weights to cancel each other out (Exp C), (3) complete agreement can boost the corresponding weights but with diminishing returns in increasing duplicates (Exp F), and (4) non-overlaps appear to be treated as a straightforward union of result sets (Exp D, Exp E).
We can also infer from these observations (and verified with additional tests) that unequal contradictions (e.g., two positives and one negative) produce a subtractive effect, even if not an exact cancellation. 

In the next section, we use these results to guide our analysis of Zeng \& Battle's dataset~\cite{Zeng2023review}, where we investigate how individual graphical perception papers induce shifts in Draco's soft constraint weights and cluster the results according to similarities in weight shift patterns.

%% file: content/5-analysis.tex
\vspace{-2mm}

\section{Exploratory Analysis: What Can We Learn From Draco Constraints?}
\label{sec:analysis}

\begin{figure*}
\centering
 \includegraphics[width=1.0\textwidth]{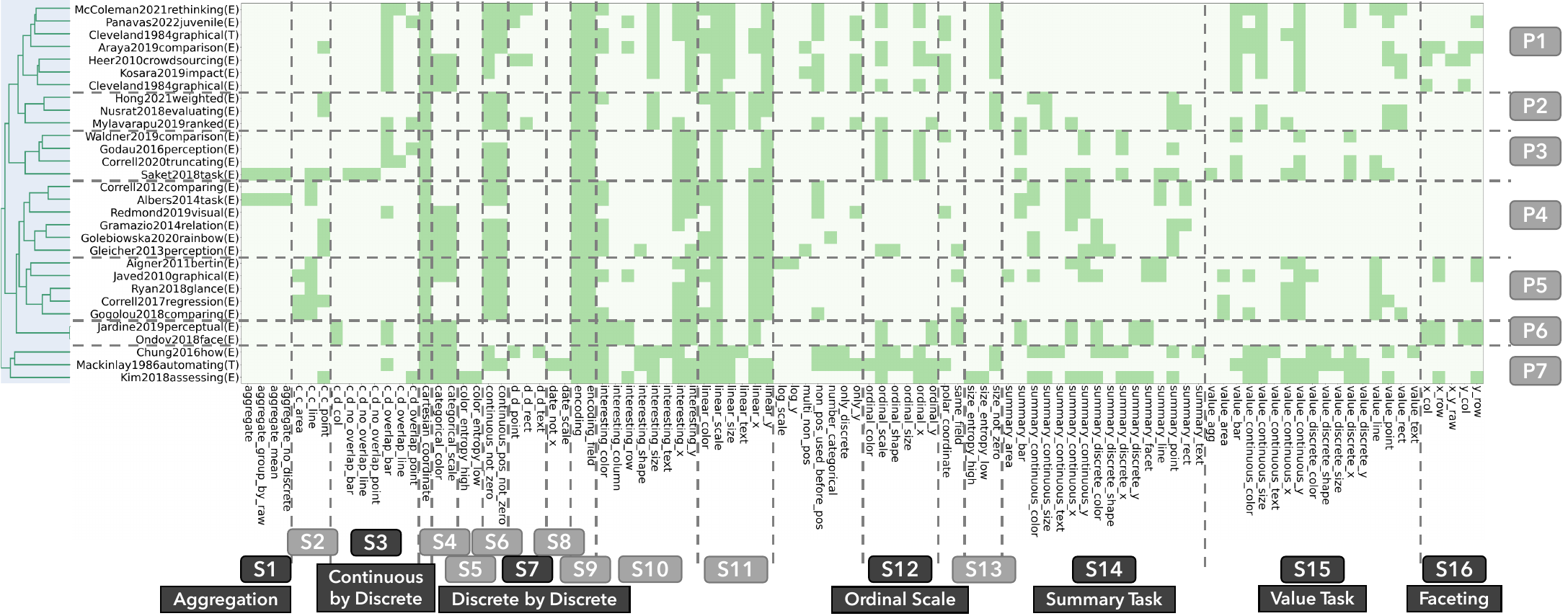}
 \vspace{-5mm}
 \caption{Clustering papers based on their covered Draco soft constraints. \vspace{-5mm}}
 \label{fig:visualization-space-coverage}
\end{figure*}

We seek to explore how different graphical perception results augment Draco's visualization recommendations, individually and collectively.
We use the following research questions to guide our exploration:
\begin{itemize}[nosep]
    \item \textbf{RQ1}: How can overlaps in visualization design coverage be quantified through Draco?
    \item \textbf{RQ2}: How might differences and similarities in Draco's soft constraint weights be used to programmatically cluster graphical perception studies?
    \item \textbf{RQ3}: To what degree does each graphical perception paper \textit{influence} Draco's recommendation behaviors?
\end{itemize}

\vspace{-1mm}

\subsection{Identifying Different Visualization Spaces}
\label{sec:analysis:visualization-space}

\begin{table}
\caption{Functionalities of Draco soft constraints covered by existing graphical perception work.}
\vspace{-3mm}
\centering
\begin{tabular}{c|l|l}
\toprule
\small \textbf{\#} & \small \textbf{Constraints} & \small \textbf{Functionality} \\
\midrule
\small \textbf{S1} & \small aggregate\_ & \small Detect whether aggregation is used \\
\rowcolor[HTML]{ededed} 
\small S2 & \small c\_c\_ & \small Detect whether scale is continuous by continuous \\
\small \textbf{S3} & \small c\_d\_ & \small Detect whether scale is continuous by discrete  \\
\rowcolor[HTML]{ededed} 
\small S4 & \small categorical\_ & \small Detect whether categorical scale is used \\
\small S5 & \small color\_ & \small Detect whether color encoding is used \\
\rowcolor[HTML]{ededed} 
\small S6 & \small continuous\_ & \small Detect whether the continuous scale includes zero baseline \\
\small \textbf{S7} & \small d\_d\_ & \small Detect whether scale is discrete by discrete \\
\rowcolor[HTML]{ededed} 
\small S8 & \small date\_ & \small Detect whether datetime field is visualized \\
\small S9 & \small encoding\_ & \small Count how many encodings are used \\
\rowcolor[HTML]{ededed} 
\small S10 & \small interesting\_ & \small Detect what encoding is used for the ``interesting'' field \\
\small S11 & \small linear\_ & \small Detect whether linear scale is used \\
\rowcolor[HTML]{ededed} 
\small \textbf{S12} & \small ordinal\_ & \small Detect whether ordinal scale is used \\
\small S13 & \small size\_ & \small Detect whether size encoding is used \\
\rowcolor[HTML]{ededed} 
\small \textbf{S14} & \small summary\_ & \small Detect whether summary task is evaluated \\
\small \textbf{S15} & \small value\_ & \small Detect whether value task is evaluated \\
\rowcolor[HTML]{ededed} 
\small \textbf{S16} & \small x\_, y\_  & \small Detect whether faceting is used \\
\bottomrule 
\end{tabular}
\label{tab:soft-constraints}
\vspace{-3mm}
\end{table}

To understand how overlaps in visualization design coverage can be detected across papers (\textbf{RQ1}), we analyze which of Draco's soft constraints are captured by each paper in our dataset.
As mentioned in \autoref{sec:pipeline}, Draco-Learn learns design preferences by inferring differences in soft constraints between positive and negative visualization examples.
To analyze the visualization design space covered by existing graphical perception work, we generate a feature vector $v$ (i.e., an embedding) for each paper.
A soft constraint $i$ is covered by a paper ($v_i = 1$) when $i$ is detected in either the positive or negative example of a ranked visualization pair from that paper.
Next, we perform a hierarchical clustering~\cite{Ward1963hierarchical,scipy.cluster} (distance function: Euclidean, linkage function: 'complete' / Farthest Point Algorithm) on papers by their feature vectors ($v$) to identify papers covering similar visualization design spaces.
The results are shown in~\autoref{fig:visualization-space-coverage}.

In total, we find \textbf{98 of 147 soft constraints are covered by the 30 graphical perception papers analyzed}. We sort Draco soft constraints alphabetically and observe that they can be separated into several groups based on their functionality, as summarized in~\autoref{tab:soft-constraints}.

According to the dendrogram on the left of \autoref{fig:visualization-space-coverage},  these papers can be clustered roughly into 7 groups, distinguished by covered experiment tasks (P1, P4, P6) and visualization design preferences (P2, P3, P5, P7).
We find that these groups are distinguished mainly by overlaps in the soft constraints they \textit{omit} rather than include.
Papers in Group P1 ignore summary tasks (S14) while those in P4 and P6 ignore value tasks (S15).
Papers in Groups P2, P3, and P4 ignore faceting visualizations (S16).
Papers in Group P5 mainly omit discrete (ordinal or nominal) scales (S3, S7, S12). 
Papers in Group P7 cover the broadest range of soft constraints and only omit aggregation-related constraints (S1).

In summary, we are able to use Draco's soft constraints to identify overlaps in the tested visualization design space of existing graphical perception papers.
For the sake of space, we only highlight a few useful observations that can be taken from this analysis.
First, our analysis enables researchers to identify visualization design decisions that may not be well covered among the papers we analyzed. For example, we observe that they cover linear scales well, followed by categorical and ordinal scales, but seldom tests log scales.
Second, this analysis enables researchers to identify new or existing perceptual tasks that have been under-evaluated with certain visualizations.
For example, we find that papers tend to focus either on value or summary tasks; only 30\% of papers (the ones in Group P2, P3, P7) evaluate visualizations in both task categories.
Thus, even when looking only at soft constraint coverage, our analysis can open up new avenues for graphical perception research, which we discuss in \autoref{subsec:discussion:benefit-studies}.

\subsection{Clustering Graphical Perception Studies by Shifts in Draco Soft Constraint Weights}
\label{sec:analysis:schools-of-thought}

Next, we explore whether groups of papers seem to shift Draco's soft constraint weights in similar or contradictory directions (\textbf{RQ2}).
In~\autoref{sec:draco-case-study}, we demonstrate how Draco reconciles agreements and conflicts under artificial scenarios.
Here, we apply the same analysis method but with real-world data, using the following model setups:
\begin{itemize}[nosep]
    \item \textbf{Baseline Model}: Draco's resulting soft constraint weights when only the APT paper~\cite{Mackinlay1986automating} is included (same as \autoref{sec:draco-case-study}).
    \item \textbf{Plus-one Models}: Draco's resulting soft constraint weights when APT~\cite{Mackinlay1986automating} and one extra paper are included, where the added paper is the focus of our analysis.
\end{itemize}
The intuition behind this approach is that graphical perception studies with similar findings should induce similar shifts in soft constraint weights.
To calculate the weight shift vector, for every soft constraint $i$, we subtract the weight of the baseline model from the plus-one model and then apply the $sign$ function to the subtraction results:
\vspace{-2mm}
\begin{equation}
    s_i = sign(p_i - b_i)
    \vspace{-2mm}
\end{equation}
where $b_i$ is the weight of soft constraint $i$ under the baseline model, $p_i$ is the weight of soft constraint $i$ under the plus-one model,
and $s_i$ indicates whether soft constraint $i$ in the plus-one model shifts positively, negatively, or remains unchanged.

\begin{figure}
\centering
 \includegraphics[width=1.0\columnwidth]{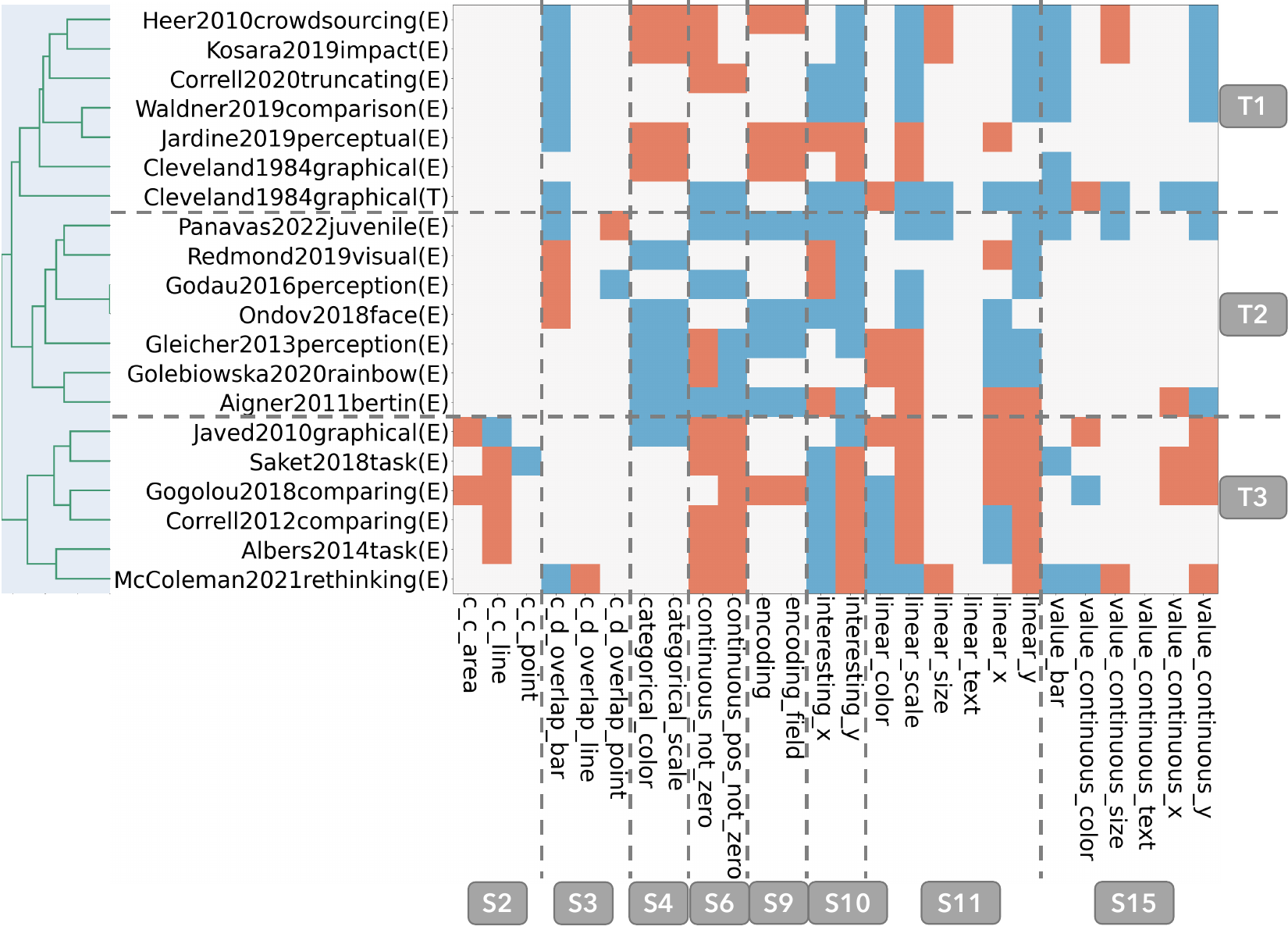}
 \vspace{-2mm}
 \caption{Clustering papers based on their resulting weight shifts in soft constraints. 
 Red represents positive shifts (higher cost), and blue negative shifts (lower cost).}
 \label{fig:schools-of-thought}
 \vspace{-6mm}
\end{figure}

To group papers by similarity, we again perform a hierarchical clustering~\cite{Ward1963hierarchical,scipy.cluster} (distance function: Euclidean, linkage function: 'complete' / Farthest Point Algorithm) on the corresponding weight shift vectors.
First, we acknowledge that this analysis did not yield definitive clusters for all 30 papers. It seems that some papers have small overlaps but no striking similarities with other papers.
That being said, we are still able to identify three major clusters spanning 20 of the 30 papers analyzed, shown in~\autoref{fig:schools-of-thought}.
To save space, we omit papers that do not fall into a cluster and only keep the related soft constraints where the remaining papers show strong visualization preferences. Note that we do not omit any significant disagreements among these papers; only soft constraints that are minimally used are filtered out.

\textbf{Papers with similar weight shifts tend to be clustered together.}
We find that within the same group, papers show similar visualization preferences. For example, Group T1 contains many papers that emphasize bar charts and linear/continuous scales and papers in Group T2 all prefer to use a color encoding for categorical scales (S4).
Between groups, we also observe strong disagreements.
For example, papers in Group T2 seem to prefer color hue encodings (S4), whereas papers in T1 disfavor color hue encodings.
As another example, papers in Group T2 favor continuous scales with a non-zero baseline (S6), while papers in T3 prefer to have a zero baseline. 
These findings suggest that clustered papers may induce a large shift on their corresponding weights when combined in Draco.

\textbf{Intra-cluster conflicts may cancel each other out in Draco.} 
It also seems that certain groupings may end up canceling each other out in Draco. For example, continuous scales may appear neutral to Draco if Groups T2 and T3 are merged when in reality there is an obvious clash in preferences among the literature. Thus, Draco's behavior may prove difficult to predict when 30 papers are integrated at once; interaction effects must be carefully considered.

\textbf{Weight shifts can be used to cluster replication/extension experiments.}
\texttt{Heer2010crowdsourcing}~\cite{Heer2010crowdsourcing} replicated \texttt{Cleveland1984graphical}~\cite{Cleveland1984graphical} to assess the use of crowdsourcing platforms for graphical perception experiments. They cover similar constraints (see P1 in \autoref{fig:visualization-space-coverage}) and also fall into the same cluster in~\autoref{fig:schools-of-thought}.
Both \texttt{Heer2010crowdsourcing} and \texttt{Cleveland1984graphical} shift soft constraints in Group S3, S4, and S9 in the same direction.
Moreover, \texttt{Heer2010crowdsourcing} also changes the weights of soft constraints in Group S11 and S15, while \texttt{Cleveland1984graphical} does not.
Upon closer investigation, we find that 
\texttt{Heer2010crowdsourcing} not only matched the results of \texttt{Cleveland1984graphical} but also added visualizations to the empirical evaluations and made all visualizations in the study comparable.
Thus, it makes sense that \texttt{Heer2010crowdsourcing} overlaps with but also exceeds \texttt{Cleveland1984graphical} in terms of shifting soft constraint weights, since \texttt{Heer2010crowdsourcing} provides similar but also greater knowledge.
These findings suggest that our approach can automatically detect significant experiment overlaps and extensions, which could help streamline the process of conducting meta-analyses and synthesizing broader graphical perception findings in the future. We discuss this further in \autoref{subsec:discussion:benefit-knowledge}.

\textbf{Papers with similar design space coverage but different weight shifts can lead to inter-cluster conflicts.}
For example, both \texttt{Jardine2019perceptual}~\cite{Jardine2020perceptual} and \texttt{Ondov2018face}~\cite{Ondov2019face} studied how people perceive small multiples with different arrangements, and these two papers cover the exact same visualization design space (Group P6 in~\autoref{fig:visualization-space-coverage}).
However, they convey different graphical perception guidelines; \texttt{Ondov2018face} shows preferences in using categorical (S4) and linear (S11) scales and visualizing interesting fields with x- or y-axes (S10). 
In contrast, \texttt{Jardine2019perceptual} shows the opposite preferences.
We find that both ranked the same set of visualizations differently since their tasks were different (aggregation vs. finding extremum), even though both tasks are categorized as summary tasks. 
For example, the stacked arrangement performed the best in \texttt{Ondov2018face} but the worst in \texttt{Jardine2019perceptual}.
What is exciting about this result is that it shows our approach can detect subtle differences between graphical perception papers with significant overlap. As an example, visualization researchers who are not graphical perception experts may miss subtle differences between \texttt{Jardine2019perceptual} and \texttt{Ondov2018face}. With our approach, these differences could be detected automatically and highlighted for non-experts, enabling them to gain a broader view of these papers.

\subsection{Quantifying Unique Influence with Shifts in Soft Constraint Weights}

When analyzing only the \textit{sign} of the weight shift, we ignore the \textit{magnitude} with which each paper shifts Draco's soft constraint weights (\textbf{RQ3}).
In this analysis, we consider both the sign and magnitude of these shifts to quantify how much a particular paper changes Draco's behavior in comparison to other papers (\autoref{sec:analysis:influence-in-constraints}). To understand how these differences manifest as practical changes in visualization recommendations, we analyze whether there is a correlation between patterns in shifting Draco's soft constraint weights and changes in Draco's actual visualization recommendations (\autoref{sec:analysis:influence-in-recommendations}).

\subsubsection{How Do Weight Shifts Map to Unique Influence?}
\label{sec:analysis:influence-in-constraints}

We reuse our baseline-plus-one setup from \autoref{sec:analysis:schools-of-thought} to calculate weight shift vectors.
However, in this case, we generate a \textit{normalized} weight shift vector for each paper to analyze its ``unique influence'', i.e., this paper's ability to change the behavior of the baseline model:
\vspace{-2mm}
\begin{equation}
    n_i = (p_i - b_i) / w_{max};
    \enspace where \enspace w_{max} = \max ( |p^1_{i}|,|p^2_{i}|, ..., |p^k_{i}|, |b_i|)
\vspace{-2mm}
\end{equation}
Here, $b_i$ is the weight of soft constraint $i$ under the baseline model, $p^k_{i}$ is the weight of soft constraint $i$ under the plus-one model $p^k$, $w_{max}$ is the maximum absolute weight of soft constraint $i$ among all models (including the baseline), and $n_i$ is the normalized difference between $b_i$ and $p_i$.
We calculate $w_{max}$ instead of $|b_i|$ as the denominator because soft constraint weights can be very small. For example, if the baseline absolute weight $|b_i|$ is close to 0, then $d_i$ would be extremely high for all plus-one models, making it hard to draw comparisons.

\begin{figure}
\centering
 \includegraphics[width=0.67\columnwidth]{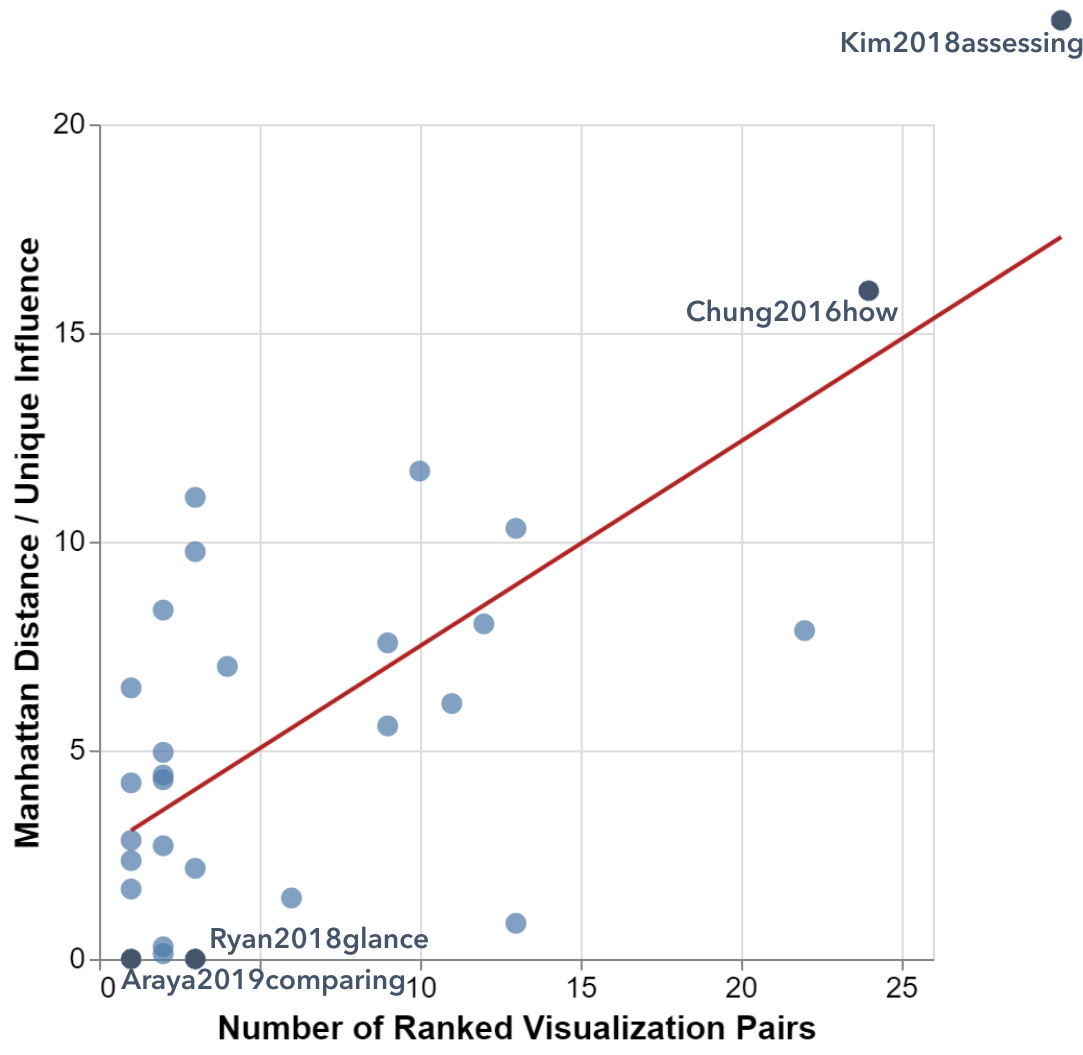}
 \vspace{-2mm}
 \caption{The relationship between the number of ranked visualization pairs a paper contributes and its ``unique influence''. We observe a moderate correlation ($r=0.5612,p<0.05$).}
 \label{fig:unique-influence-scatter}
 \vspace{-6mm}
\end{figure}
To measure a paper's ``unique influence'' over the baseline model, we calculate the Manhattan distance (i.e., the $L^1$-norm) of its weight shift vector.
Then, we plot the papers in a scatterplot with the number of ranked visualization pairs associated with each paper on the x-axis and the Manhattan distance on the y-axis, as shown in~\autoref{fig:unique-influence-scatter}.
We notice that some papers such as \texttt{Kim2018assessing}~\cite{Kim2018assessing} and \texttt{Chung2016how}~\cite{Chung2016how} exert a significant influence on Draco's soft constraint weights, i.e., significantly change the behavior of the baseline. Meanwhile, other papers such as \texttt{Ryan2018}~\cite{Ryan2018glance} and \texttt{Araya2019comparison}~\cite{Araya2019comparision} seem to exert ``zero'' influence, i.e., do not shift any of the soft constraint weights in the baseline model.

First, we investigate why two experimental results~\cite{Araya2019comparision,Ryan2018glance} do not change any of Draco's soft constraint weights.
According to our mini-experiments in \autoref{sec:draco-case-study}, zero weight shifts in all soft constraints suggest that adding these particular papers may not provide any ``new information'' when going from the baseline model to the plus-one models. 
We observe two possible explanations for zero shifts on the soft constraint weights: (1) the default soft constraints from Draco fail to cover the differences between the ranked pairs of visualization designs tested in the target study, i.e., Draco does not sufficiently support the kinds of constraints discussed in the corresponding paper; or (2) the target study provides similar graphical perception knowledge as the baseline model, i.e., these papers suggest redundant knowledge. 

\textbf{In some cases, Draco's soft constraints fail to cover the differences between ranked visualizations.}
Upon closer examination of these results, we find that Draco's default soft constraints fail to cover the difference between the positive and negative visualization designs extracted from the studies of \texttt{Ryan2018}, aligning with our first explanation.
Further, we find that Draco's soft constraints cannot detect the difference between two visualization designs with the same encoding and mark setting but different field statistics (e.g., cardinality, entropy, etc.). 
\texttt{Ryan2018} evaluated the performance of line charts visualizing data with different entropy, where Draco fails to detect the entropy difference within ranked visualization pairs.

\textbf{Providing visualization design knowledge already covered by the baseline could result in ``zero'' unique influence.} Unlike \texttt{Ryan2018}, we verify that Draco is, in fact, able to discern differences between the positive and negative examples extracted for \texttt{Araya2019comparison}.
Compared to the visualization preferences of the baseline, we find that the soft constraints that are only seen in the positive example from \texttt{Araya2019comparison}, such as \mintinline{prolog}{preference(linear_x, E)}, \mintinline{prolog}{preference(linear_size, E)} and \mintinline{prolog}{preference(value_continuous_size, E)}, already have negative weights in the baseline model.
Similarly, soft constraints that are only found in the negative example from \texttt{Araya2019comparison}, such as \mintinline{prolog}{preference(value_bar, M)} and \mintinline{prolog}{preference(ordinal_scale, E)}, are already disfavored by the baseline model.
In other words, Draco can discern and prioritize differences among these soft constraints with identical accuracy using the APT~\cite{Mackinlay1986automating} baseline dataset. Thus, the added data from \texttt{Araya2019comparison} appears to be redundant and therefore yields a model identical to the baseline.

\textbf{Providing ``more'' visualization comparison knowledge could boost unique influence.}
As we investigate the reasons why a graphical perception work could have ``zero'' unique influence, we also consider the opposite: under what circumstances does a paper rank highly in terms of unique influence?
First, we see that the top two ``most influential'' papers, \texttt{Kim2018assessing} and \texttt{Chung2016how}, are in Group P7 in~\autoref{fig:visualization-space-coverage}, which covers the broadest range of soft constraints.
There are two ways in which graphical perception papers can increase coverage of soft constraints: either they provide \textit{more} positive-negative visualization pairs (i.e., test many pairs) and/or they provide \textit{greater diversity} of visualization pairs (e.g., test many unique visualization pairs).
To test this, we consider the number of ranked visualization pairs that each graphical perception paper provides and observe a notable correlation in \autoref{fig:unique-influence-scatter}.
Furthermore, we find that 10 graphical perception papers provide more than five ranked visualization pairs, and 7 of these 10 papers rank within the top 10 most ``influential'' papers in~\autoref{fig:unique-influence-scatter}. These findings suggest that most papers boost their unique influence by yielding more pairs of positive and negative examples that Draco can train on. That being said, we also observe papers that have relatively fewer total pairs but still rank highly in unique influence, suggesting that greater diversity of pairs also plays an important role. We posit that maximizing both measures (number and diversity of paired visualizations) can further boost a paper's unique influence measure. We discuss these opportunities in \autoref{subsec:discussion:benefit-knowledge}.

\vspace{-2mm}
\subsubsection{How Do Shifts in Soft Constraint Weights Correlate with Shifts in Recommendations?}
\label{sec:analysis:influence-in-recommendations}

\begin{figure}
\centering
 \includegraphics[width=0.6\columnwidth]{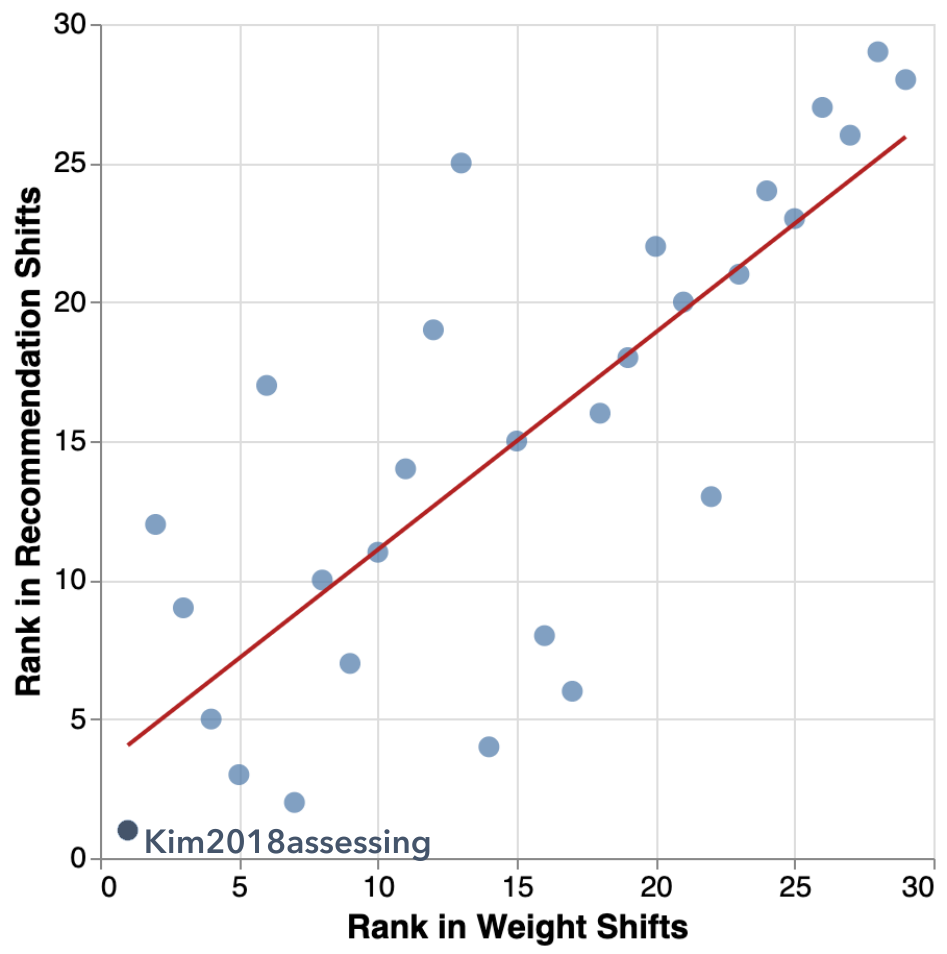}
 \vspace{-2mm}
 \caption{We see a strong linear correlation($r=0.7818, p<0.001$) between shifts in soft constraint weights (x-axis) and shifts in visualization recommendation rankings (y-axis).}
 \vspace{-6mm}
 \label{fig:recommendation-shifts}
\end{figure}

Although Draco's soft constraint weights are a convenient structure for our analysis, it is unclear how shifts in soft constraint weights translate to shifts in actual recommendations.
In response, we investigate how shifts in Draco's soft constraints affect the visualization recommendations that it generates.
Inspired by the benchmarking approach of Lin et al.~\cite{Lin2020dziban}, we analyze shifts in Draco's recommendations when visualizing three datasets: IMDB movies~\cite{vega_movies} (3201 rows),
Cars~\cite{vega_cars} (406 rows),
and Seattle Weather~\cite{vega_seattle_weather} (1461 rows).
Various combinations of fields are also visualized, e.g., Nominal $\times$ Quantitative, Quantitative $\times$ Quantitative, Nominal $\times$ Quantitative $\times$ Quantitative, etc. We evaluate recommendation shifts using 461 total combinations of attributes.

To evaluate shifts in recommendations, we follow a similar approach to our weight-shift analysis, modified slightly to account for the many visualizations generated by Draco.
Since Draco has its own default weights for soft constraints (hand-tuned by the Draco authors), we request this default Draco to generate its top 100 recommendations based on each combination of data attributes.
Then, we ask both the baseline and plus-one models to rank these 100 visualizations, generating a corresponding rank vector $r$ where $|r| = 100$.
The value of a cell $r^\textit{baseline}_i$ in the baseline rank vector $r^\textit{baseline}$ is the rank of how the baseline favors the $i$-th recommendation from the default Draco.

To measure a paper's recommendation shifts, we calculate the Spearman Rank Correlation between the baseline and plus-one rank vectors, producing 461 Spearman correlation values for each paper (one per attribute combination).
We calculate the average of these values to produce a final score for each paper.
A paper with a higher Spearman Correlation indicates that its recommendations are aligned with the baseline's recommendations, meaning fewer and/or smaller recommendation shifts.
We plot each paper's rank in recommendation shifts against its rank in unique influence (i.e., weight shifts) in~\autoref{fig:recommendation-shifts}.
We find a strong correlation between these two ranks ($r=0.7818, p<0.001$).
This result indicates that large shifts in Draco's soft constraint weights also translate into significant shifts in visualization recommendation rankings. Thus, Draco's soft constraint weights can be a useful proxy for measuring differences in recommendation behaviors.

%% file: content/6-discussion.tex
\vspace{-1mm}
\section{Discussion}
\label{sec:discussion}

In this paper, we use Draco as a probe to explore how agreements and disagreements among the graphical perception literature can translate into shifts in the behavior of visualization recommendation algorithms. In this section, we show how our findings can lead to new research directions and discuss the limitations of our analysis.

\vspace{-1mm}
\subsection{Benefits to Perception \& Recommendation Research}

\subsubsection{Implication: Normalizing Current Knowledge}
\label{subsec:discussion:benefit-knowledge}

Our results show that aligned graphical perception studies induce similar shifts in Draco's soft constraints, whereas opposed studies may cancel each other out.
For example, consider a conflict on the influence of perception bias on position encodings~\cite{Gogolou2018comparing,Xiong2019biased}. Analyzing weight shifts on Draco's soft constraints could provide a quantitative means of reconciling these observed discrepancies, e.g., by measuring the degree to which these papers may cancel each other out in Draco's soft constraint weights.
Our results also suggest a correlation between shifts in Draco's soft constraint weights and changes in its generated visualization recommendations. 
As a result, Draco's soft constraint weights could provide an \textit{intermediate representation} for graphical perception results, enabling researchers to quantitatively compare many different empirical and theoretical findings.

Broadly, our research introduces the concept of \emph{quantitative meta-analysis} in graphical perception, an established methodology in medicine for integrating findings from multiple, individual experiments that have yet to be adopted in computer science~\cite{sutton2000methods}. Future research can build on this concept to derive hypothesis-driven meta-analyses for graphical perception; for example, by translating existing methods from medicine for managing heterogeneity across experiment designs.

\subsubsection{Application: Contributing New Knowledge}
\label{subsec:discussion:benefit-studies}

Given a common denominator by which to compare graphical perception studies, our analysis approach also reveals an opportunity to quantify the \textit{influence} of current and future graphical perception work. For example, our analysis approach provides a means of measuring overlap between a particular study compared to the prior literature. This approach can benefit the graphical perception community in two ways. First, it could help the community surface well-covered areas of graphical perception knowledge; in other words, areas that may not require further replication studies, highlighted by the corresponding soft constraints. For example, according to~\autoref{fig:visualization-space-coverage}, soft constraints \texttt{linear\_y}, \texttt{linear\_color} and \texttt{linear\_size} (\textit{using y/color/size encoding for linear scale}) are already well discussed in the literature.

Second, our approach can help to identify knowledge gaps among a given set of papers. For example, our findings show that some Draco constraints are under-utilized by the 30 papers we analyzed, such as the use of logarithmic scales (\texttt{log\_y} and \texttt{log\_scale}). To detect gaps that could be filled by new studies, this approach could be repeated by importing a targeted set of studies into Draco, such as all studies testing the perception of log scales, using the schema by Zeng \& Battle (see \autoref{fig:zeng-battle-exp}) and our Draco mapping code as guides. Then, our methods in \autoref{sec:analysis} can help readers extract constraints with no coverage or contradictory weight shifts. These constraints represent a partial ASP specification, which can be completed by (a) manually adding constraints (e.g., adding missing data type or encoding constraints by hand) or (b) passing the partial ASP specification to Draco's recommendation engine. Then, the completed specifications can be passed to Draco's renderer to generate input visualizations for new graphical perception studies (e.g., visualizations to evaluate log scales).

\subsubsection{Reflection: Improving Recommendation Frameworks}
\label{subsec:discussion:benefit-framework}

Several visualization recommendation frameworks have been proposed to make it easier to develop new recommendation algorithms~\cite{Wongsuphasawat2016towards,Moritz2018formalizing,Siddiqui2017fast}.
To the best of our knowledge, this paper is the \textit{first} to incorporate a large body (i.e., 30 papers) of graphical perception results into a visualization recommendation framework. In our analysis, we find that 
only 30 of the 59 papers in the Zeng \& Battle dataset could be translated for use with Draco (see \autoref{sec:pipeline:zeng-battle}).
Further, we observe that other frameworks fail to support large-scale integration. For example, Zenvisage/ZQL~\cite{Siddiqui2017fast} does not support perceptual rankings. CompassQL~\cite{Wongsuphasawat2016towards} only allows users to select one ranking method and requires user-defined functions to express perceptual rankings beyond the default.
Even when using Draco, the default Draco soft constraints cannot capture the difference between all positive and negative visualization examples in some cases (see \autoref{sec:analysis:influence-in-constraints}).
These findings suggest that some, and in some cases, all graphical perception results cannot be easily expressed by existing visualization recommendation frameworks.
By revealing their limitations, our work highlights opportunities to make visualization recommendation frameworks more expressive and aligned with current graphical perception work.
For example, future frameworks could follow a SQL language structure like ZQL/CompassQL while also supporting multiple simultaneous ranking strategies like Draco.

\subsubsection{Evolution: Automated Updates to Recommenders}
\label{sec:discussion:benefit:updates}

Finally, our research contributes an initial pipeline for mapping graphical perception results directly into input data for Draco. In the future, this work could be extended to support the full range of Zeng \& Battle's JSON specification, which would enable the automated translation of graphical perception results into Draco models. In this way, if new graphical perception papers are added to Zeng \& Battle's dataset in the future, automated pipelines could allow them to be ingested by any tool that uses Draco to generate recommendations, similar to the concept of semantic versioning in software package development.
This approach could also reduce the need for visualization tool designers to manually identify and reconcile conflicts in existing graphical perception guidelines, as Draco could handle them automatically. That being said, our findings in \autoref{sec:analysis} show how it may be unwise to blindly aggregate graphical perception studies to produce quick results.
For example, automated reconciliation may mask serious inter-cluster and intra-cluster study conflicts and must be done with care (see \autoref{sec:analysis:schools-of-thought}). Further, naive approaches to integrating these studies may cause smaller studies to be drowned out by larger ones, but bigger is not necessarily better (see \autoref{sec:analysis:influence-in-constraints}). We need better techniques for integrating large numbers of studies into visualization recommendation systems, which requires tighter collaboration between graphical perception and visualization recommendation researchers moving forward.

\vspace{-1mm}
\subsection{Limitations \& Future Work}

\subsubsection{Graphical Perception Findings Evolve Over Time.}
\label{subsec:discussion:limitation-dataset}

We stress that our analyses represent a synthesis of a limited body of work at a specific point in time. Further, they are not verbatim quotes from the original authors; others are welcome to augment or even present alternative interpretations of these works.
Several of the papers we analyze are also exploratory in nature and do not necessarily draw explicit conclusions, e.g., \cite{McColeman2021rethinking,Panavas2022juvenile},
causing us to shy away from making stronger claims than the original authors themselves might allow.
As new research is added or Draco is updated, the exact values in the results will shift; however, the focus of this work is on the \emph{overall pipeline} rather than specific results.
This contribution is strengthened further by the ease with which our analyses can be updated to incorporate new advancements in graphical perception, as mentioned in \autoref{sec:discussion:benefit:updates}.

\subsubsection{Limitations in Draco Influence Our Results}
\label{subsec:discussion:limitation-framework}

As we mentioned in \autoref{sec:pipeline:draco-limitations}, \autoref{sec:analysis:influence-in-constraints} and \autoref{subsec:discussion:benefit-framework}, the current Draco soft constraints fail to identify differences between a small fraction of ranked visualization pairs.
One could extend Draco's soft constraints to provide better coverage of the Zeng \& Battle dataset~\cite{Zeng2023review} and other graphical perception studies in the future.
Our findings may also reveal a lack of communication between graphical perception and visualization recommendation researchers that could be improved. Further, more granular study variables such as selected color palettes are omitted from this analysis, which could explain some of our findings. More research is needed to capture finer-grained variables in future taxonomies to support quantitative meta-analysis.

Another possible extension is to develop applications to support researchers. For example, providing tools for analyzing the weight shifts and recommendation shifts more deeply to investigate whether the resulting weights and recommendations truly align with the theoretical and empirical results in current graphical perception work.